\documentclass{webofc}
\usepackage[varg]{txfonts}   

\usepackage{multicol}
\usepackage{wrapfig}

\usepackage{color}
\definecolor{darkred}{rgb}{0.4,0.0,0.0}
\definecolor{darkgreen}{rgb}{0.0,0.4,0.0}
\definecolor{darkblue}{rgb}{0.0,0.0,0.4}

\newcommand{\be}{\begin{equation}}
\newcommand{\ee}{\end{equation}}
\newcommand{\beq}{\begin{equation}}
\newcommand{\eeq}{\end{equation}}
\newcommand{\bea}{\begin{eqnarray}}
\newcommand{\eea}{\end{eqnarray}}

\newcommand{\bie}{\begin{small} \begin{itemize}}

\begin{document}
\title{QCD vacuum replicas are metastable}
%
%

\author{\firstname{Pedro} \lastname{Bicudo}\inst{1}\fnsep\thanks{\email{bicudo@tecnico.ulisboa.pt}} \and
        \firstname{José Emílio} \lastname{Ribeiro}\inst{1}\fnsep\thanks{\email{emilioribeiro@tecnico.ulisboa.pt}} 
}

\institute{
CeFEMA, Departamento de Física, Instituto Superior Técnico, Universidade de Lisboa.
          }

\abstract{%
We study the meson spectra in the excited QCD vacua, denominated replicas.
We find all mesons have real masses, with no tachyons, thus showing the QCD replicas are indeed metastable.

}
\maketitle
%

%
%
\section{Introduction}

\subsection{Spontaneous breaking of chiral symmetry and the QCD vacua}

In 1960 Gell-Mann and Levy, with the sigma model, and soon after Nambu and Jona-Lasinio modelled for the first time the spontaneous breaking of chiral symmetry. The sigma model is 62 years old, what have we learned since then?

This is an effective $\phi^4$ model: there are two extrema,
 the particle masses$^2$ are the vertical curvature, 
 in the true vacuum  ${m_\sigma}^2 > 0, \ {m_\pi}^2=0$,
 in the false vacuum  ${m_\sigma}^2 < 0,  \ {m_\pi}^2<0$ we have tachyons since it is an unstable vacuum,
and it is not renormalizable.


The sigma model advanced physics a lot and is still used in Higgs sector of the standard model. However it is just a model, it is not a renormalizable field theory.

In particle physics we have a theory with spontaneous breaking of chiral symmetry: QCD, as verified by lattice QCD simulations. However we don't yet know how to study the QCD vacuum with the lattice.

In this talk I will show theoretical evidence that the QCD vacuum has a much richer structure than the sigma mode, with a tower of metastable vacua. These are the extrema we expect for QCD. We denominate these novel vacua the replicas of QCD.


\subsection{Approach of Gaussian approximation to QCD with two cumulants}

Our approach is QCD in the Gaussian approximation for two cumulants  \cite{Dosch:1988ha,Kalashnikova:2005tr}. This includes for instance the case of quark models, of the Dyson-Schwinger approach at the rainbow-ladder level of truncation and of the large $N_c$ limit of QCD,
\begin{eqnarray}
&&H=\int\, d^3x \left[ \psi^{\dag}( x) \;(m_0\beta -i{\vec{\alpha}
\cdot \vec{\nabla}} )\;\psi( x)\;+
{ 1\over 2} \int d^4y\, \
\right.
\nonumber \\
&&
\overline{\psi}( x)
\gamma^\mu{\lambda^a \over 2}\psi ( x)  
g^2 \langle A_\mu^a(x) A_\nu^b(y) \rangle
\;\overline{\psi}( y)
\gamma^\nu{\lambda^b \over 2}
 \psi( y)  \ + \ \cdots
\label{hamilt}
\end{eqnarray}

With this two cumulant diagram we can compute the dressed quark mass, the vacuum stability, boundstates, the decay of resonances, etc. These effects account for 98 \% of the nucleon mass.

The quark kernel, $K^{ab}_{\mu\nu}(x,y)$,  is derived from gluon configurations \cite{Bicudo:1998bz,Szczepaniak:2001rg}  it can be represented \cite{Bicudo:1989sh,Bicudo:1989si,Bicudo:2003cy,Bicudo:2010qp} in 
Eq.(\ref{hamilt}) as,
\be
K^{ab}_{\mu\nu}(x,y)=g^2 \langle A_\mu^a(x) A_\nu^b(y) \rangle.
=\delta^{ab} \Gamma_{\mu \nu} K_0^{\alpha+1}|\vec{x}-\vec{y}|^\alpha.
\label{potenciais}
\ee

From all these studies, it emerged that although numerical values may vary somewhat, as for instance when one goes from harmonic to linear confining kernels, the global overall physical  picture does not. 

Importantly, all the PCAC  theorems, such as the Gell-Mann Oakes and Renner relation for $M_\pi$ and $\langle \bar \psi \psi \rangle$, the Goldberger and Treiman relation and the Weinberg Theorem for $\pi \pi$ scattering, the Alder zero for pion couplings are verified \cite{LeYaouanc:1984ntu,llanesEstrada:2003ha,Bicudo:2003fp}.

Moreover this generates a constituent mass for the light quark, leading to chiral symmetry restoration for excited hadrons, while for heavy quarks, a quasi classical regime for hadrons must set in \cite{Glozman:2005tq}.


In Eq. (\ref{hamilt} ),  $m_0$ is  the current mass of the quark, for simplicity we will set $m_0$ to zero.

%
\begin{figure}[t]
\includegraphics[width=0.7\columnwidth,trim=60pt 60pt 0 20pt ]{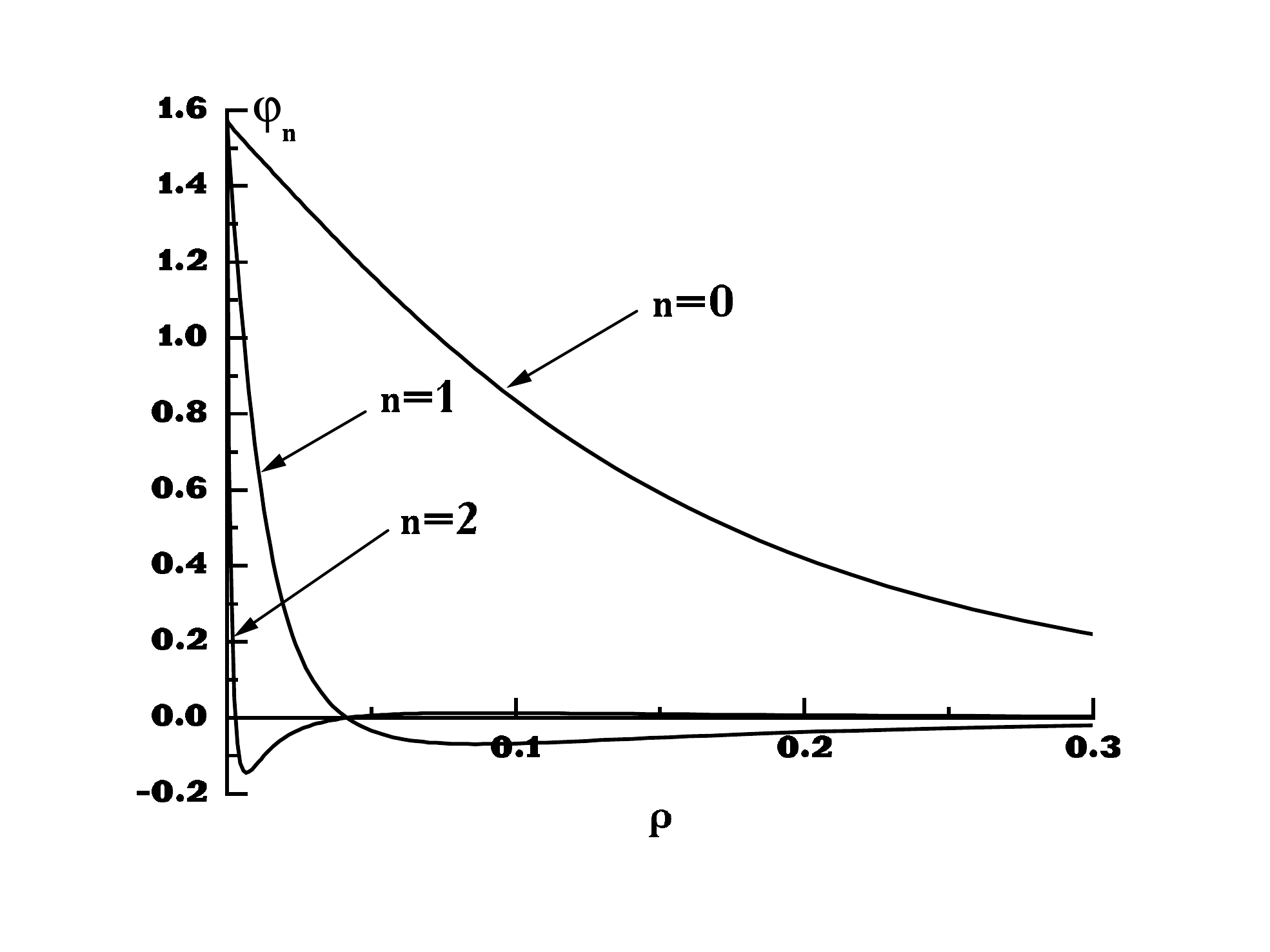}
\caption{
The first solution, $\varphi_0(k)$, of the mass gap equation, representing the vacuum $| 0 \rangle$ and its two first two other solution, $\varphi(k)_1$ and $\varphi(k)_2$ for a linear confining potential \cite{Bicudo:2003cy,Bicudo:2010qp}, corresponding to the truncated Coulomb gauge of QCD), computed in the chiral limit of $m_{q\,\;bar{q}}=0$.
}\label{fig:replicaslinear}
\end{figure}

The big picture is independent of the details of the kernel: all the results will apply equally well with any confining potential. In this talk and for numerical simplicity, we will use the harmonic confining kernel.

The model of Eq.(\ref{hamilt}) was suggested in the mid-eighties \cite{Amer:1983qa,LeYaouanc:1984ntu,LeYaouanc:1983huv}, and latter re-derived in terms of coherent states of $^3P_0$ quark-antiquark pairs \cite{Bicudo:1989sh,Bicudo:1989si}.

\subsection{Our previous knowledge of the replicas}


%
\begin{figure}[t]
\includegraphics[width=0.50\columnwidth]{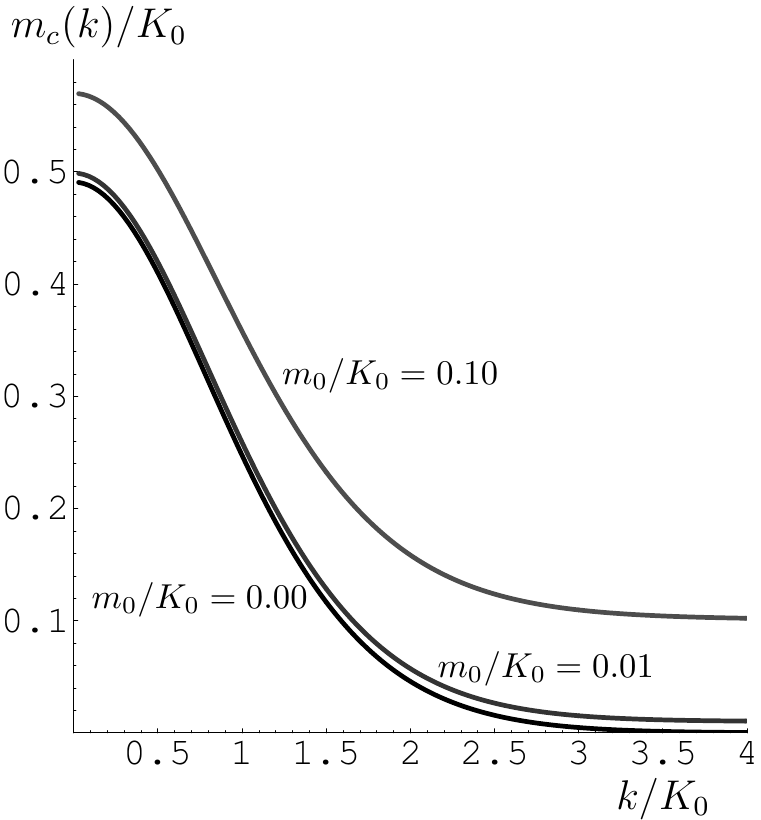}
\caption{
The dependence \cite{Bicudo:2006dn} of the constituent quark mass $m_c(k)$ on the current quark mass $m_0$.
}\label{fig:masssolution}
\end{figure}

With confining kernels, there is an infinite tower of excited vacua-like states, the replicas $| {\cal R}1 \rangle$ , interpolating between the true chiral symmetry breaking, physical vacuum, to the highest chiral invariant vacuum \cite{LeYaouanc:1984ntu,Bicudo:1989sh,Bicudo:2003cy}. 
Examples of solutions in the case of the linear potential \cite{Bicudo:2003cy} for the chiral angle $\varphi$ 
are depicted in Fig. \ref{fig:replicaslinear}.  

The mass gap equation for the vacua has similarities with the Bethe Salpeter equation for Hadrons in the true vacuum, who also have an infinite tower of excited states. This is a robust evidence for the replicas.


We also know that the number of tachyon states for the false chiral invariant vacuum becomes infinite 
\cite{Bicudo:2006dn} in the pseudoscalar and scalar channels,
whereas, from the true vacuum, we have an infinite set of true hadronic states. This shows the false vacuum is unstable while the true vacuum $| 0 \rangle$ is stable, it is an absolute minimum.

For infinite volumes, the different vacua correspond to orthogonal $^3P_0$ coherent states so that, two different coherent states,  have a zero overlap between them, hence  being independent of each other. In addition, they are separated apart by an infinite amount of energy
\cite{Bicudo:2002eu,AntNefRib-2010} . 
The energy density is $({\cal E}_{{\cal R}1}\simeq  87.2476\;MeV/fm^3$ (using $K_0\simeq 300 MeV$). For instance, for a bubble with finite a radius of only $5$ fm, we would had to add to any hadronic excitation in ${\cal R}1$ another $45$ GeV, and this for an overlap $\langle {\cal R}1_| 0 \rangle =2.9 \times 10^{-16}$.


%
%
\section{Details of the mass gap and bound state equations}

\subsection{Mass gap equation and replicas}

There are three equivalent methods to solve the mass gap equation:
\begin{enumerate}
\item assume a quark-antiquark $^3P_0$ 
coherent state, and minimize its correspondent energy density,
\item
rotate the quark fields by a Bogoliubov-Valatin canonical transformation to diagonalize the Hamiltonian,
\item
solve the Schwinger-Dyson equations for the quark propagators. 
\end{enumerate}

Any of these methods
lead to the same mass gap equation and $m_c(k)$.

Technically, it is convenient to decompose the relativistic invariant Dirac-Feynman propagators
\cite{LeYaouanc:1984ntu,Bicudo:1998mc}, 
in the quark and antiquark Bethe-Goldstone 
propagators, with quark and antiquark spinors $u_s$ and $v_s$, close to the formalism of non-relativistic quark models,
\begin{eqnarray}
{\cal S}_{Dirac}(k_0,\vec{k})
={i \over \not k -m +i \epsilon}= {\sum_su_su^{\dagger}_s \beta \over k_0 -E(k) +i \epsilon} 
- {\sum_sv_sv^{\dagger}_s \beta  \over -k_0 -E(k) +i \epsilon}
\label{eq:propagators}
\end{eqnarray}

Replacing the propagator
of Eq. (\ref{eq:propagators}) in its Schwinger-Dyson equation, 
{\small \begin{eqnarray}
\label{2 eqs}
&&0 = u_s^\dagger(k) \left\{k \widehat k \cdot \vec \alpha + m_0 \beta
-\int {d w' \over 2 \pi} {d^3k' \over (2\pi)^3}
i V(k-k') \right.
\nonumber \\
&&\left. \sum_{s'} \left[ { u(k')_{s'}u^{\dagger}(k')_{s'} 
 \over w'-E(k') +i\epsilon}
-{ v(k')_{s'}v^{\dagger}(k')_{s'} 
  \over -w'-E(k')+i\epsilon} \right]
\right\} v_{s''}(k) \  \
\nonumber \\
&&E(k) = u_s^\dagger(k) \left\{k \widehat k \cdot \vec \alpha + m_0 \beta
-\int {d w' \over 2 \pi} {d^3k' \over (2\pi)^3}
i V(k-k')  \right.
\nonumber \\
&&\left. \sum_{s'} \left[ { u(k')_{s'}u^{\dagger}(k')_{s'} 
 \over w'-E(k') +i\epsilon}
-{   v(k')_{s'}v^{\dagger}(k')_{s'}  
 \over -w'-E(k')+i\epsilon} \right]
\right\} u_s(k),
\end{eqnarray}        }

It was shown that for the wide class of confining potentials \cite{Bicudo:2003cy}, with the exponent $ 0<\alpha \leq 2 $ -- see Eq.(\ref{potenciais}) -- it leads to an infinite number of solutions of the mass gap equation.

The two cases mostly studied in the literature are the quadratic case, $\alpha=2$, derived in the Gaussian approximation to QCD and using the Balitsky Local Coordinate Gauge \cite{Bicudo:1998bz}, and the linear case $\alpha=1$, derived in the Coulomb gauge. 
In the case of a linear confining potential, we get integral equations. They are finite but need a regularization of infrared divergences.

With the simple harmonic interaction
\cite{LeYaouanc:1984ntu}, 
the integral of the potential is a Laplacian, and the mass gap equation can be transformed into a differential equation. The mass gap equation and the quark energy are finally given by,
{\small \begin{eqnarray}
\label{mass gap}
&&\Delta \varphi(k) = 2 k S(k) -2 m_0 C(k) - { 2 S(k) C(k) \over k^2 } \nonumber  \\ 
&&E(k) = k C(k) + m_0 S(k) - { {\varphi'(k) }^2 \over 2 } - { C(k)^2 \over k^2 }  \ .
\end{eqnarray}          }

Numerically, this equation is a non-linear ordinary differential
equation. It can be solved with the Runge-Kutta and shooting method.

Examples of solutions,
for different light current quark masses $m_0$, 
are depicted in Fig. \ref{fig:masssolution}. The effect of a small finite current quark mass $m_0 \sim 0.01 K_0 $, typical of the light $u$ and $d$ quarks, can be easily estimated as a small increase of the dynamically generated constituent quark mass $m_c$ and does not concern us here.

Examples of replica solutions, for the linear potential, are shown in Fig. \ref{fig:replicaslinear}.


\subsection{Salpeter-RPA equations for mesons}

%
%
\begin{table}[h]
\centering
\caption{
The positive and negative energy spin-independent, spin-spin, spin-orbit and 
tensor potentials are shown.
$\varphi'(k)$, ${\cal C}(k)$ and ${\cal G}(k)= 1 - S(k) $ are functions of the 
quark mass $m_c(k)$.
}
\label{spindependent} 
\begin{tabular}{c|c}
& $V^{++}=V^{--}$  \\ \hline
spin-indep. & $- {d^2 \over dk^2 } + { {\bf L}^2 \over k^2 } + 
{1 \over 4} \left( {\varphi'_q}^2 + {\varphi'_{\bar q}}^2 \right) 
+ {  1 \over k^2} \left( {\cal G}_q +{\cal G}_{\bar q}  \right) $  \\ 
spin-spin & $ {4 \over 3 k^2} {\cal G}_q {\cal G}_{\bar q} {\bf S}_q \cdot {\bf S}_{\bar q} $  \\ 
spin-orbit & $ {1 \over  k^2} \left[ \left( {\cal G}_q + 
{\cal G}_{\bar q} \right) \left( {\bf S}_q +{\bf S}_{\bar q}\right) 
+\left( {\cal G}_q - {\cal G}_{\bar q} \right) \left( {\bf S}_q -{\bf S}_{\bar q}\right)  \right]
\cdot {\bf L} $  \\ 
tensor & $ -{2 \over  k^2} {\cal G}_q {\cal G}_{\bar q} 
\left[ ({\bf S}_q \cdot \hat k ) ({\bf S}_{\bar q} \cdot \hat k )
-{1 \over 3} {\bf S}_q \cdot {\bf S}_{\bar q} \right] $ \\ \hline
& $V^{+-}=V^{-+}$ \\ \hline
spin-indep. & $0$  \\ 
spin-spin & $ -{4 \over 3} \left[ {1\over 2} {\varphi'_q} {\varphi'_{\bar q}} + 
{1\over k^2} {\cal C}_q {\cal C}_{\bar q}  \right]
{\bf S}_q \cdot {\bf S}_{\bar q} $  \\ 
spin-orbit & $0$  \\ 
tensor & $ \left[ -2 {\varphi'_q} {\varphi'_{\bar q}} + 
{2\over k^2} {\cal C}_q {\cal C}_{\bar q}  \right]
\left[ ({\bf S}_q \cdot \hat k ) ({\bf S}_{\bar q} \cdot \hat k )
-{1 \over 3} {\bf S}_q \cdot {\bf S}_{\bar q} \right] $
\end{tabular}
\end{table}

The Salpeter-RPA equations for a meson (a colour singlet
quark-antiquark bound state) can be derived either from the $q\bar{q}$ Lippman-Schwinger
equations, or by replacing the propagator
of Eq.(\ref{eq:propagators}) in the Bethe-Salpeter equation. In either way, one gets
\cite{Bicudo:1998mc},
\begin{eqnarray}
\chi(k,P) =
\int {d^3k' \over (2\pi)^3} V(k-k') \left[ 
u(k'_1)\phi^+(k',P)v^\dagger(k'_2)  +v(k'_1){\phi^-}^t(k',P) u^\dagger(k'_2)\right] 
\nonumber \\
\phi^+(k,P) = { u^\dagger(k_1) \chi(k,P)  v(k_2) 
\over +M(P)-E(k_1)-E(k_2) }
\ , \
{\phi^-}^t(k,P) = { v^\dagger(k_1) \chi(k,P) u(k_2)
\over -M(P)-E(k_1)-E(k_2)} \ .
 \label{homo sal}
\end{eqnarray}

These equations 
\cite{Bicudo:1989si,LlanesEstrada:1999uh} have positive 
$\phi^+$ and negative $\phi^-$ energy wave functions. 
The Pauli $\vec \sigma$ matrices in the spinors of Eq.(\ref{eq:propagators}), 
produce the 
\cite{Bicudo:1991kz} 
potentials of Table \ref{spindependent}.


%
%
\subsection{Numerical solution of the mass gap and boundstate equations.}

Notice that both the pseudoscalar and scalar equations
have a system with two equations.  However, the spin-dependent interactions 
couple an extra pair of equations, both in the vector and axial-vector channels.
We now combine the spin-dependent potentials of Table \ref{spindependent},
to derive the full Salpeter-RPA radial bound state
equations. 

\begin{figure}[h]
\includegraphics[width=0.30\columnwidth]{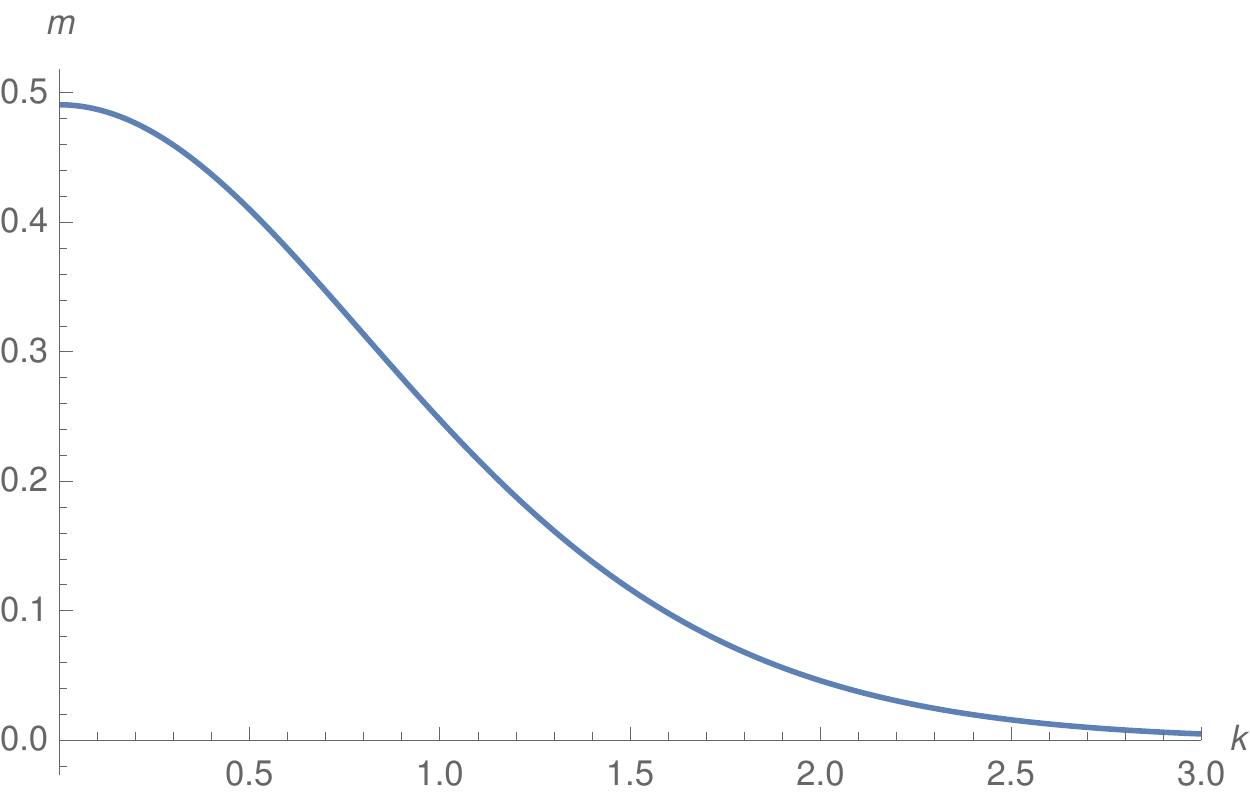}
\includegraphics[width=0.30\columnwidth]{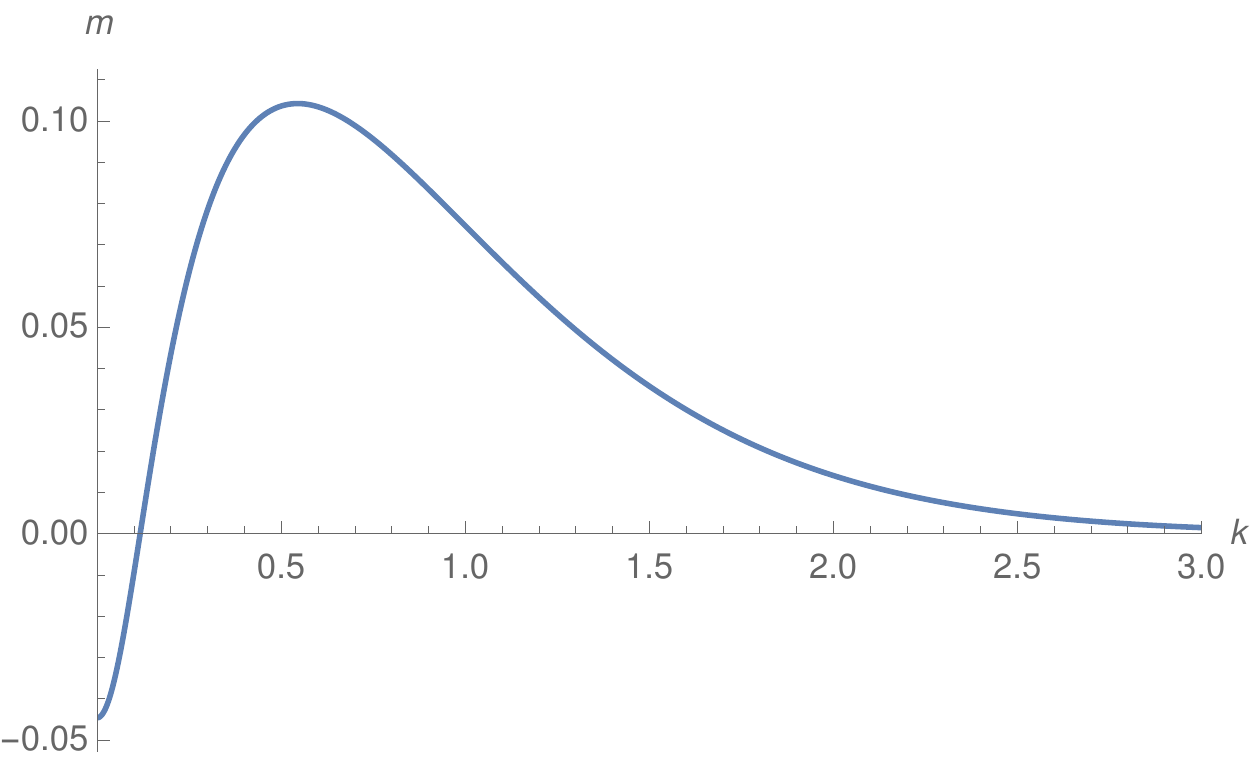}
\includegraphics[width=0.30\columnwidth]{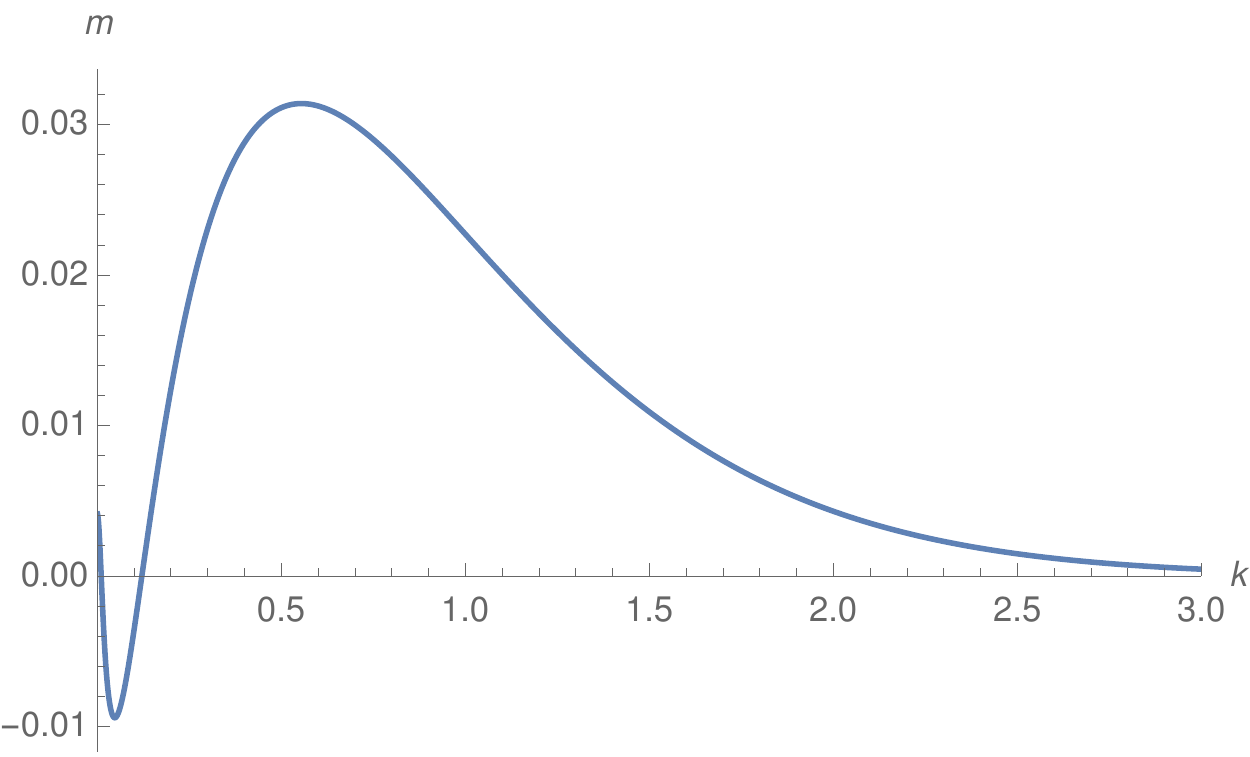}
\caption{
The constituent quark masses $m_c(k)$ in the chiral limit $m_0=0$, solutions of the mass gap equation,
from left to right for the ground state vacuum and first two replicas. 
}\label{fig:masssolutionII}
\end{figure}

\begin{figure}[h]
\includegraphics[width=0.30\columnwidth]{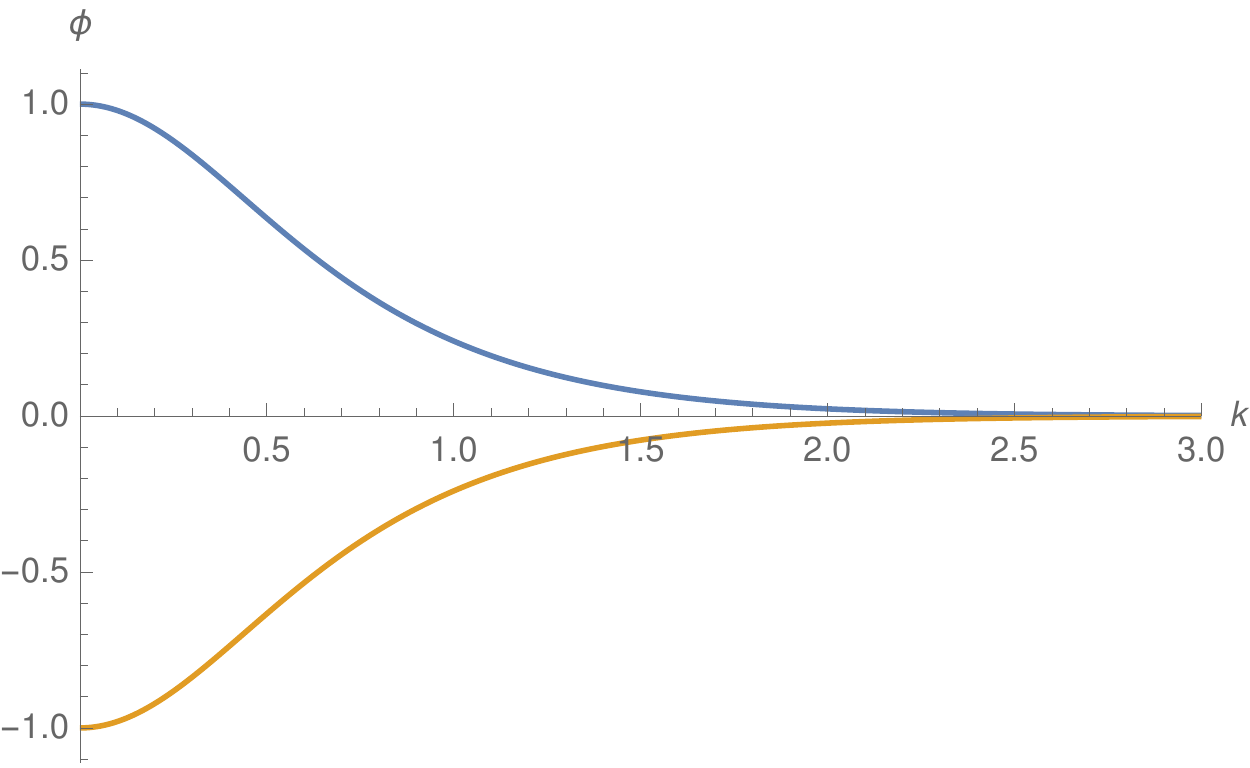} \hspace{1pt}
\includegraphics[width=0.30\columnwidth]{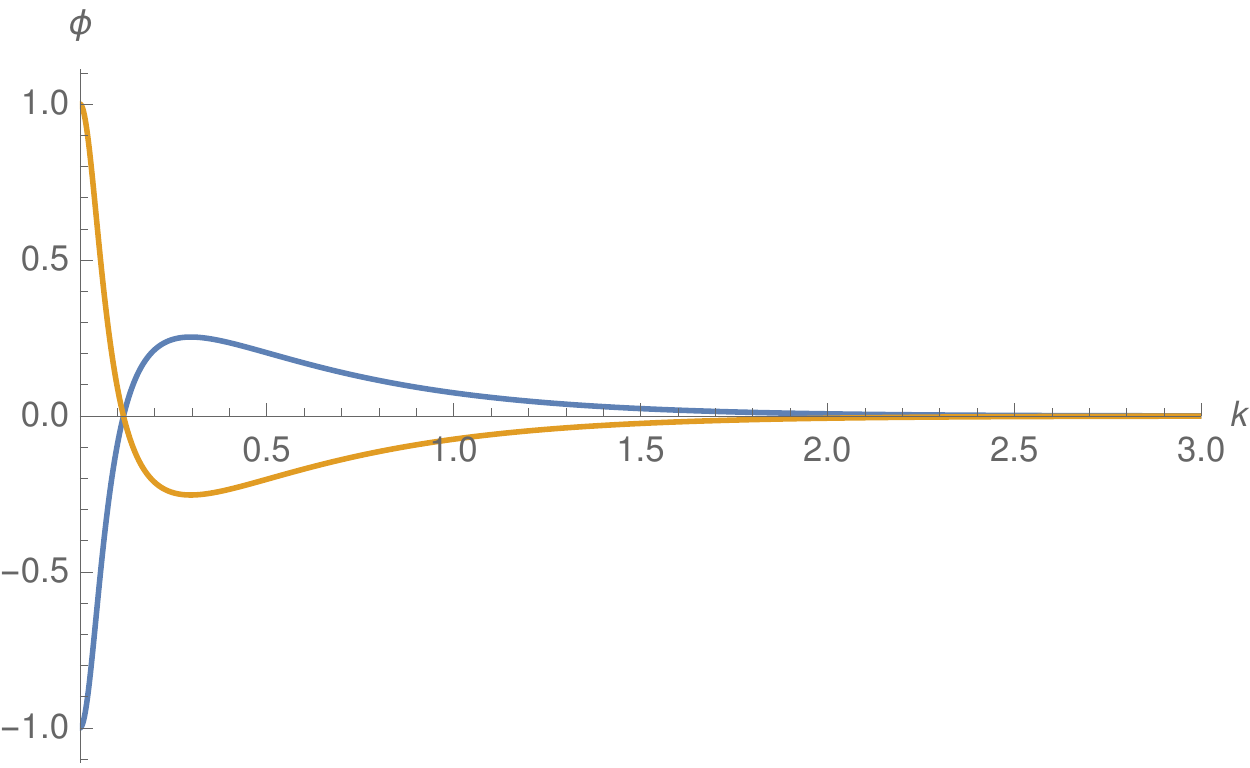} \hspace{1pt}
\includegraphics[width=0.30\columnwidth]{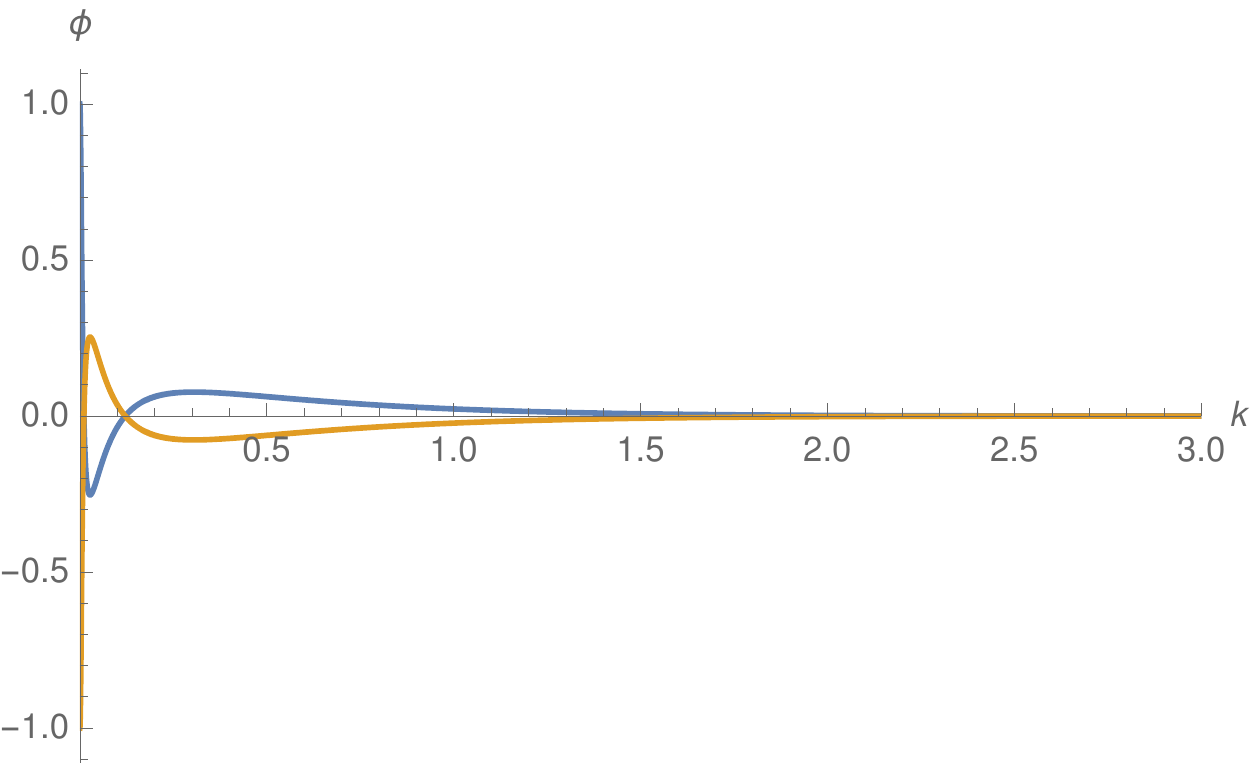}
\caption{The $J^{PC}=0^{-+}$,  $^1 S_0$ normalized pseudoscalar (P) radial wave functions  $\phi^+$ (in blue) and $\phi^-$ (in yellow), from left to right for the ground state vacuum and first two replicas, in dimensionless in units of $K_0=1$. Because the normalization diverges in the chiral limit, we arbitrarily normalize the wave functions with $\phi(0)=1$. Importantly, the wavefunctions $\phi^\pm$ are identical to $\pm \sin \varphi(k) =\pm m_c(k) /\sqrt{ k+ m_c(k)}$ as is easily verified from Fig. \ref{fig:masssolution}.}
\label{fig:pseudoscalar}
\end{figure}


\begin{figure}[h]
\includegraphics[width=0.30\columnwidth]{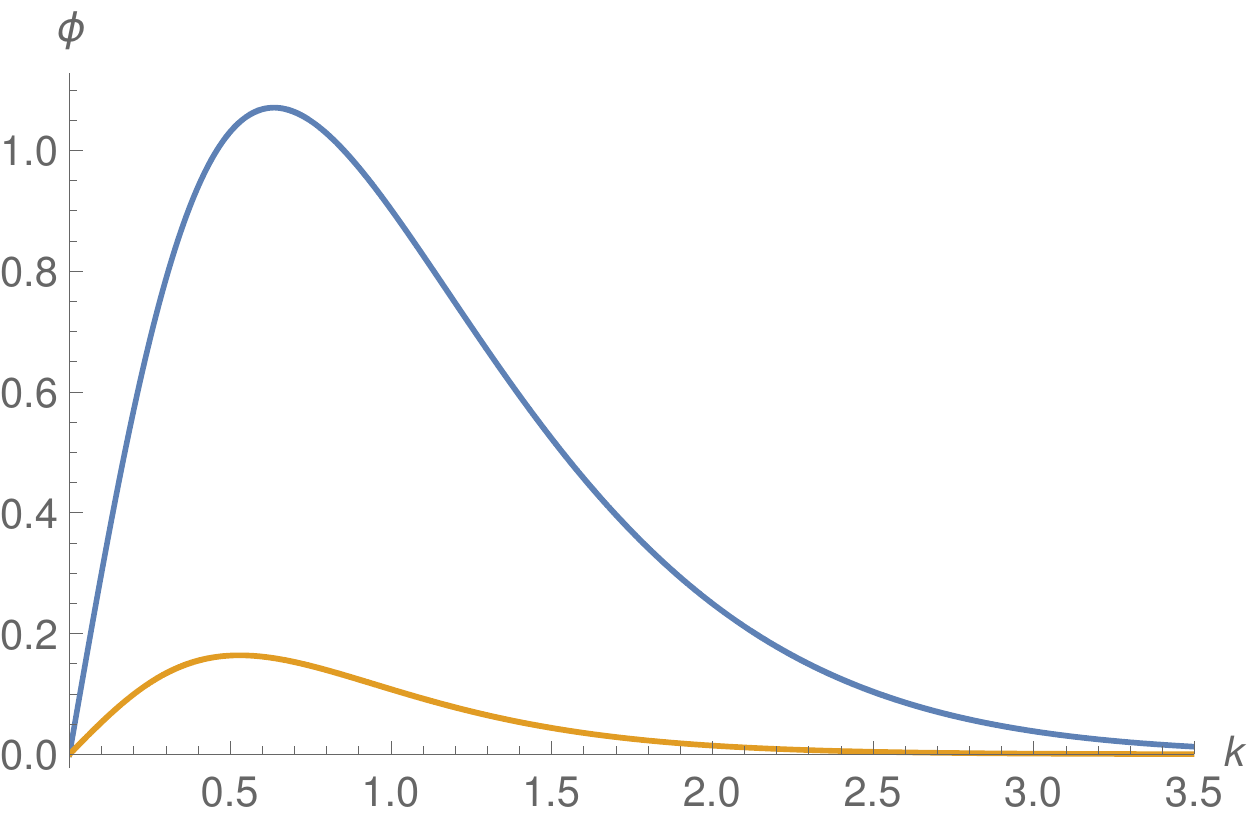} \hspace{1pt}
\includegraphics[width=0.30\columnwidth]{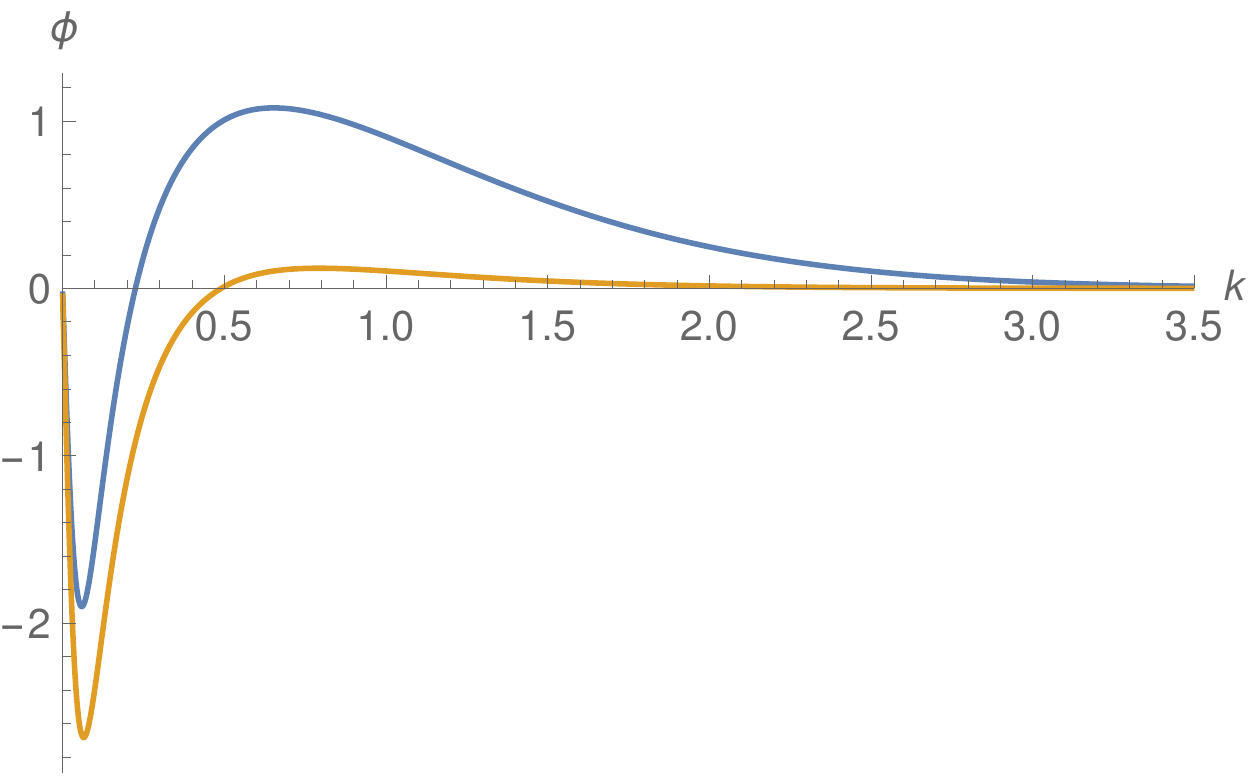} \hspace{1pt}
\includegraphics[width=0.30\columnwidth]{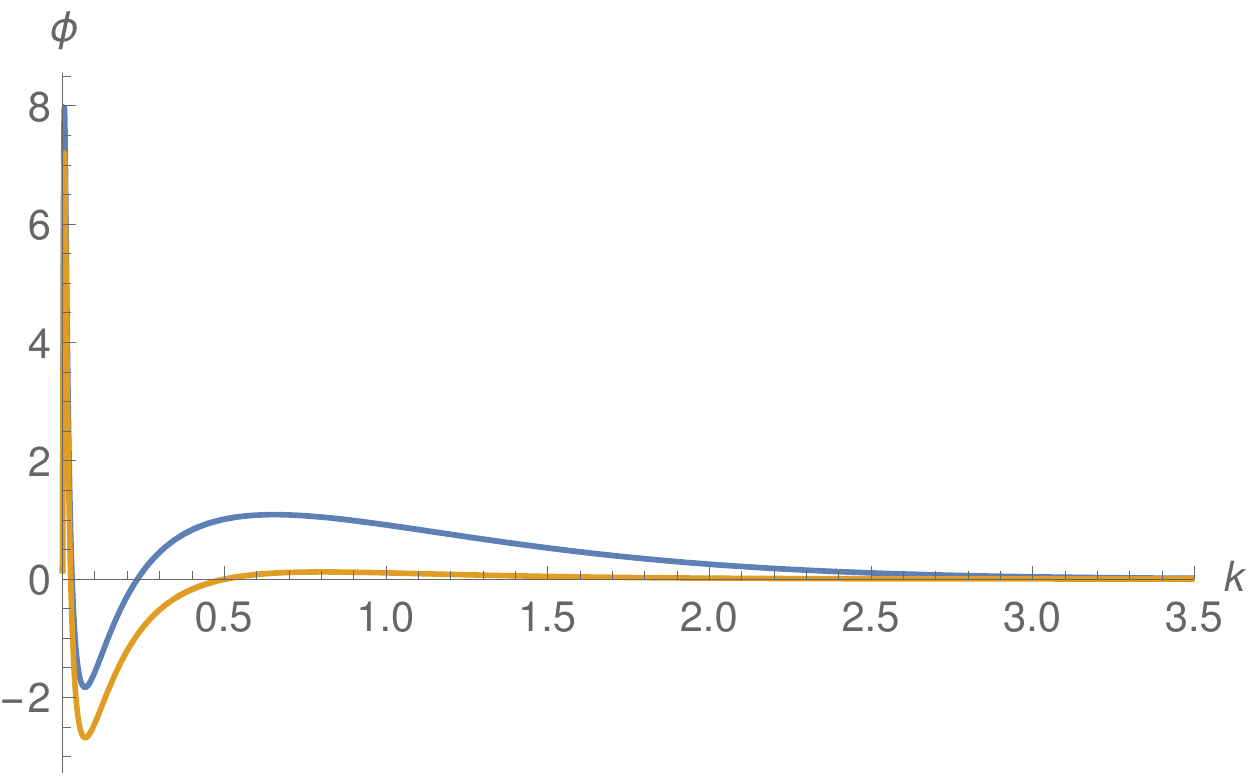}
\caption{The $J^{PC}=0^{++}$, $^3 P_0$ normalized scalar (S) radial wave functions $\phi^+$ (in blue) and $\phi^-$ (in yellow), from left to right for the ground state vacuum and first two replicas, in dimensionless in units of $K_0=1$. 
}\label{fig:scalar}
\end{figure}
\begin{figure}[h]
\includegraphics[width=0.30\columnwidth]{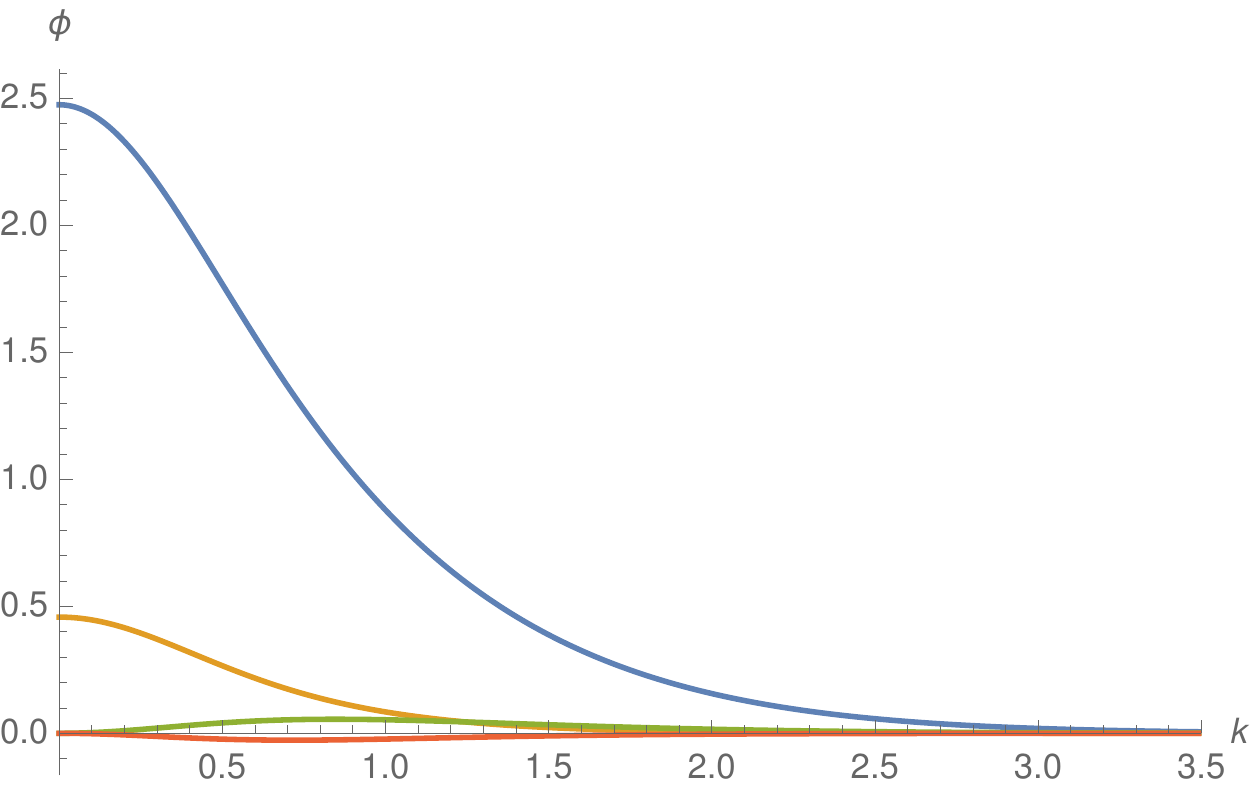} \hspace{1pt}
\includegraphics[width=0.30\columnwidth]{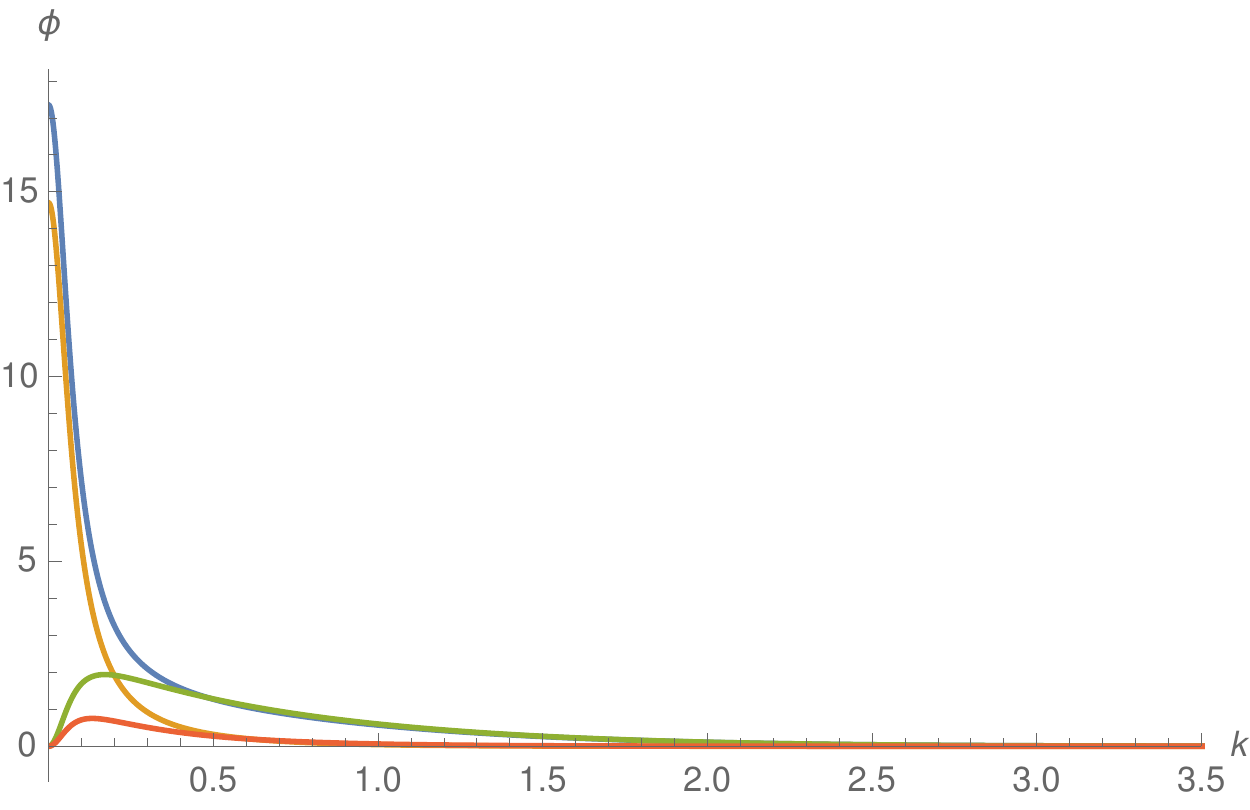} \hspace{1pt}
\includegraphics[width=0.30\columnwidth]{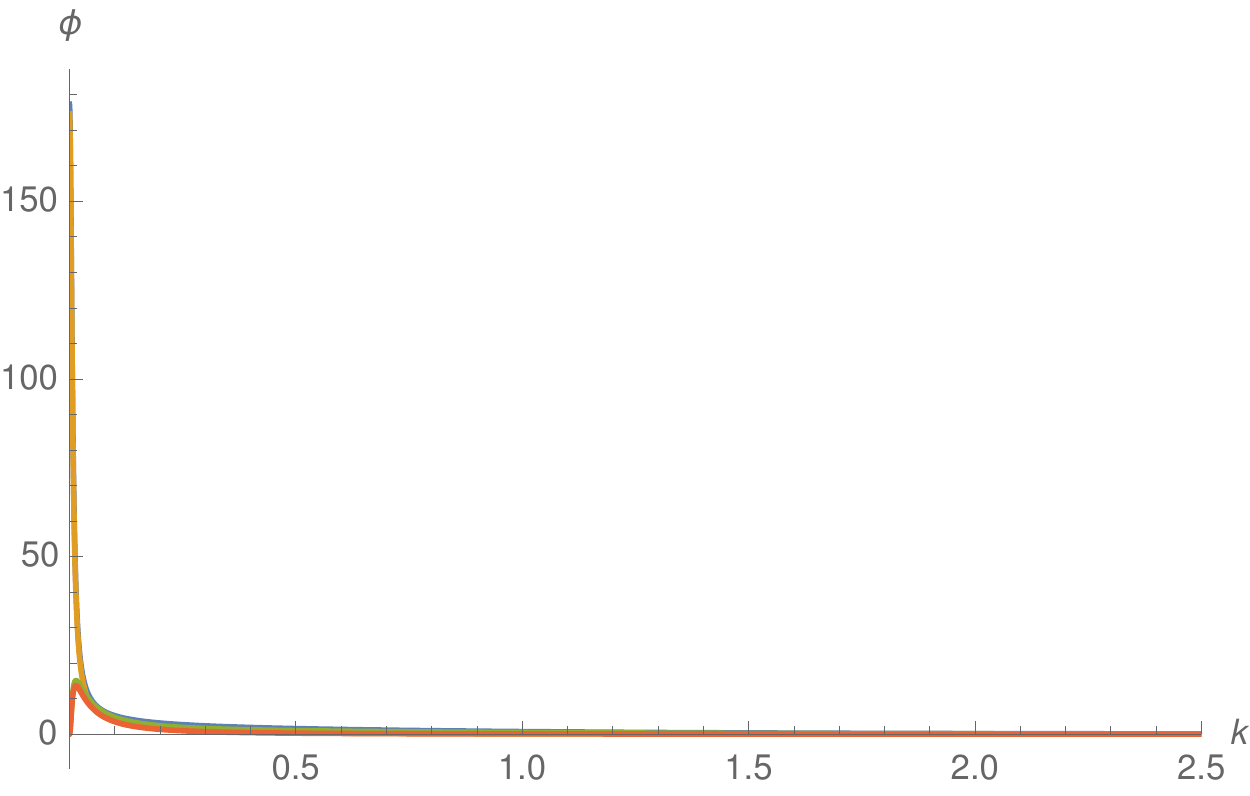}
\caption{The $J^{PC}=1^{--}$, $^3 P_1$ normalized vector (V) radial wave functions $\phi_0^+$ (in blue), $\phi_0^-$ (in yellow), $\phi_2^+$ (in green) and $\phi_2^-$ (in red) from left to right for the ground state vacuum and first two replicas,  in dimensionless in units of $K_0=1$. 
}\label{fig:vector}
\end{figure}

To solve numerically our differential equations, we use finite centred differences for the Laplacian. Since our mass gap equation is non-linear, we utilize the shooting method, so as to have the mass to vanish at a large enough UV momentum cutoff $K$ . We use a very fine mesh in momentum space because the replicas have nodes, quite close to the momentum origin. 


\begin{figure}[h]
\includegraphics[width=0.30\columnwidth]{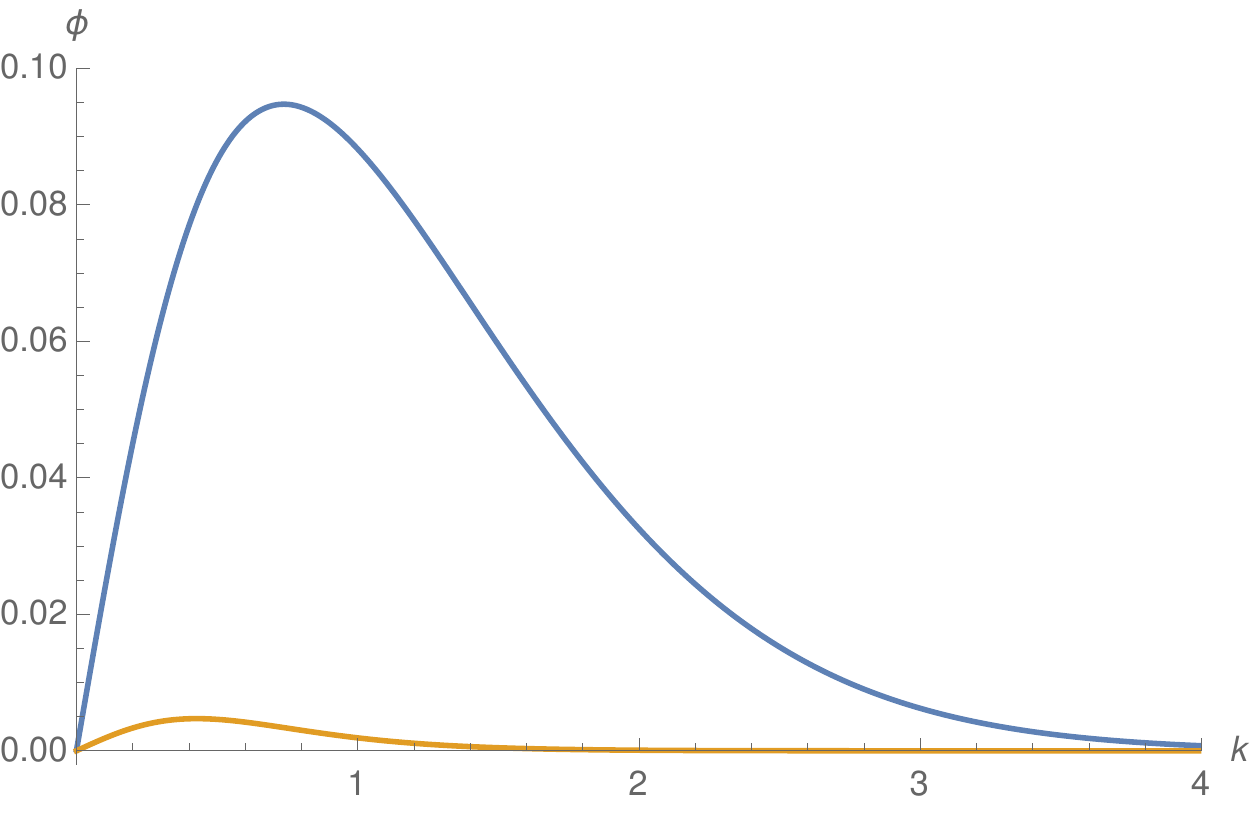} \hspace{1pt}
\includegraphics[width=0.30\columnwidth]{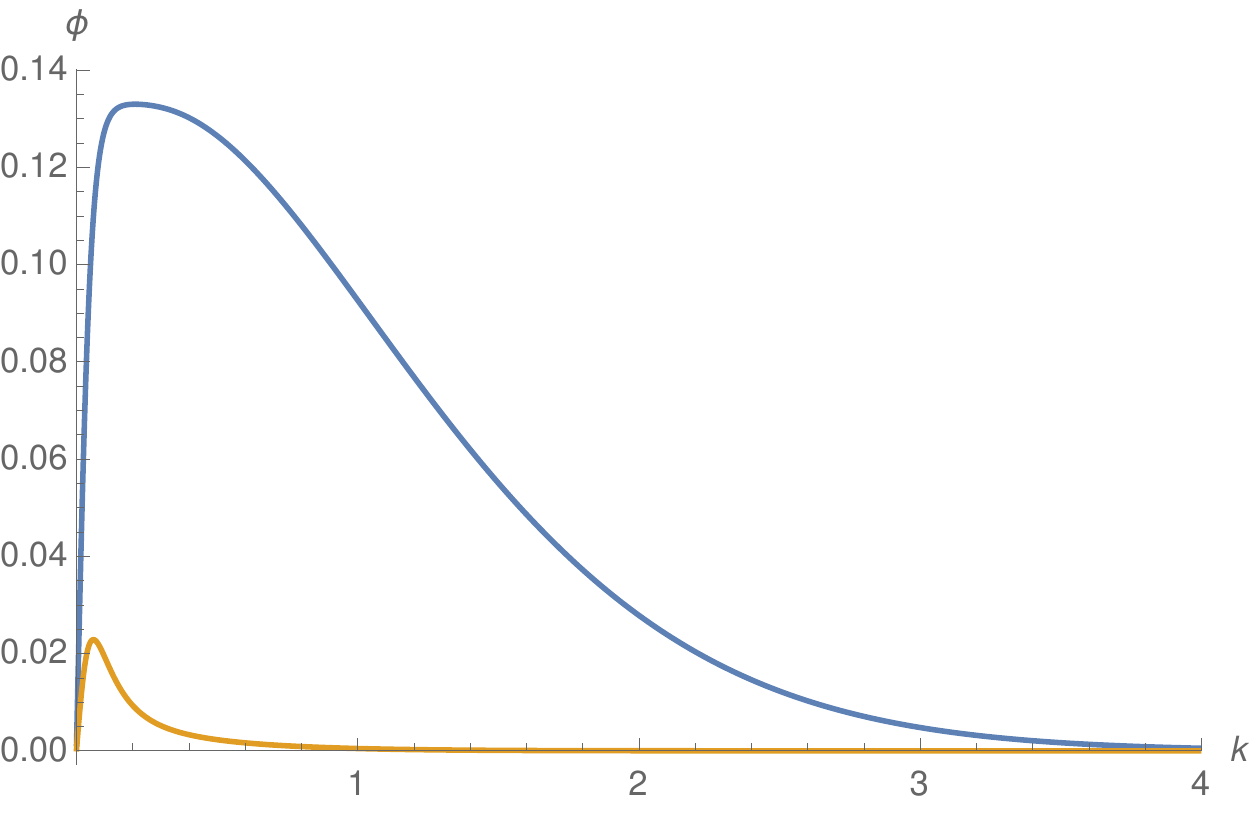} \hspace{1pt}
\includegraphics[width=0.30\columnwidth]{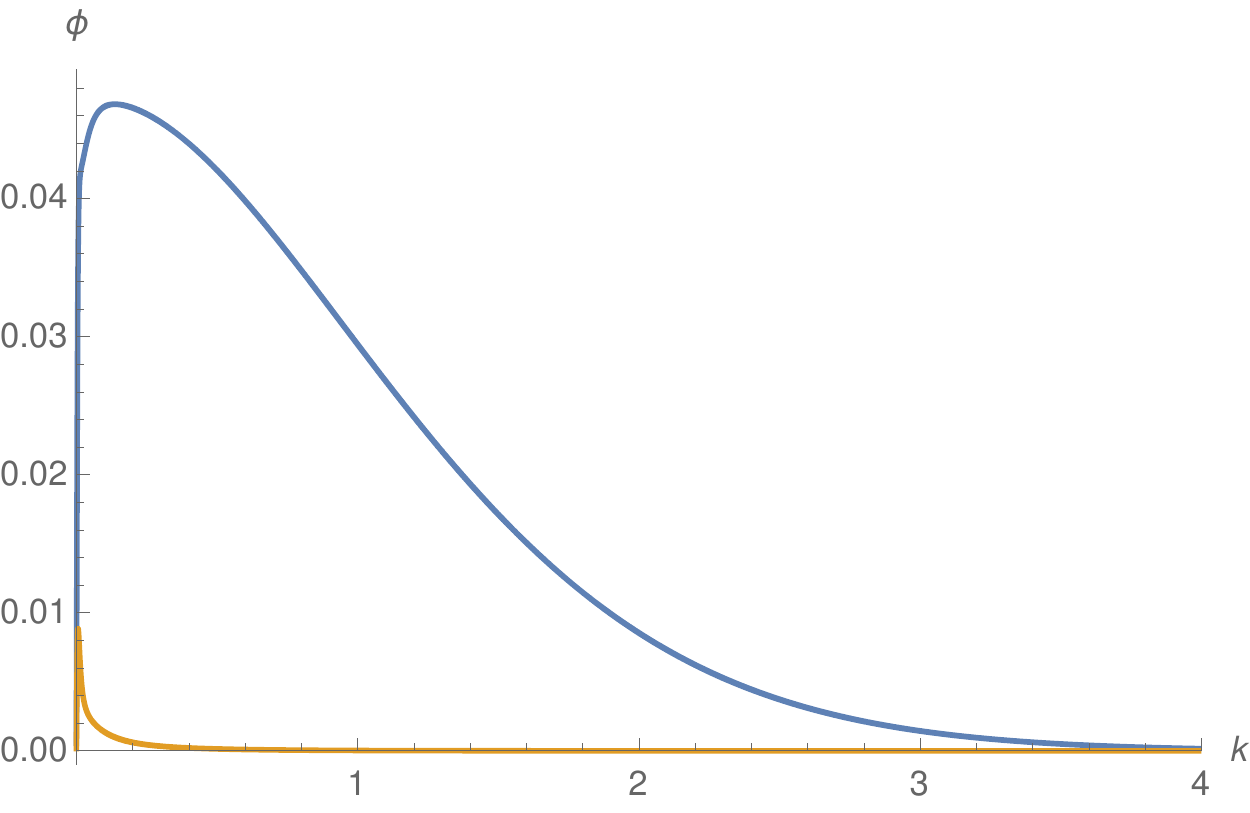}
\caption{The $J^{PC}=1^{+-}$, $^1 P_1$ normalized axial vector (A) radial wave functions $\phi^+$ (in blue) and $\phi^-$ (in yellow), from left to right for the ground state vacuum and first two replicas, in dimensionless in units of $K_0=1$. 
}\label{fig:axialvector1}
\end{figure}

\begin{figure}[h]
\includegraphics[width=0.30\columnwidth]{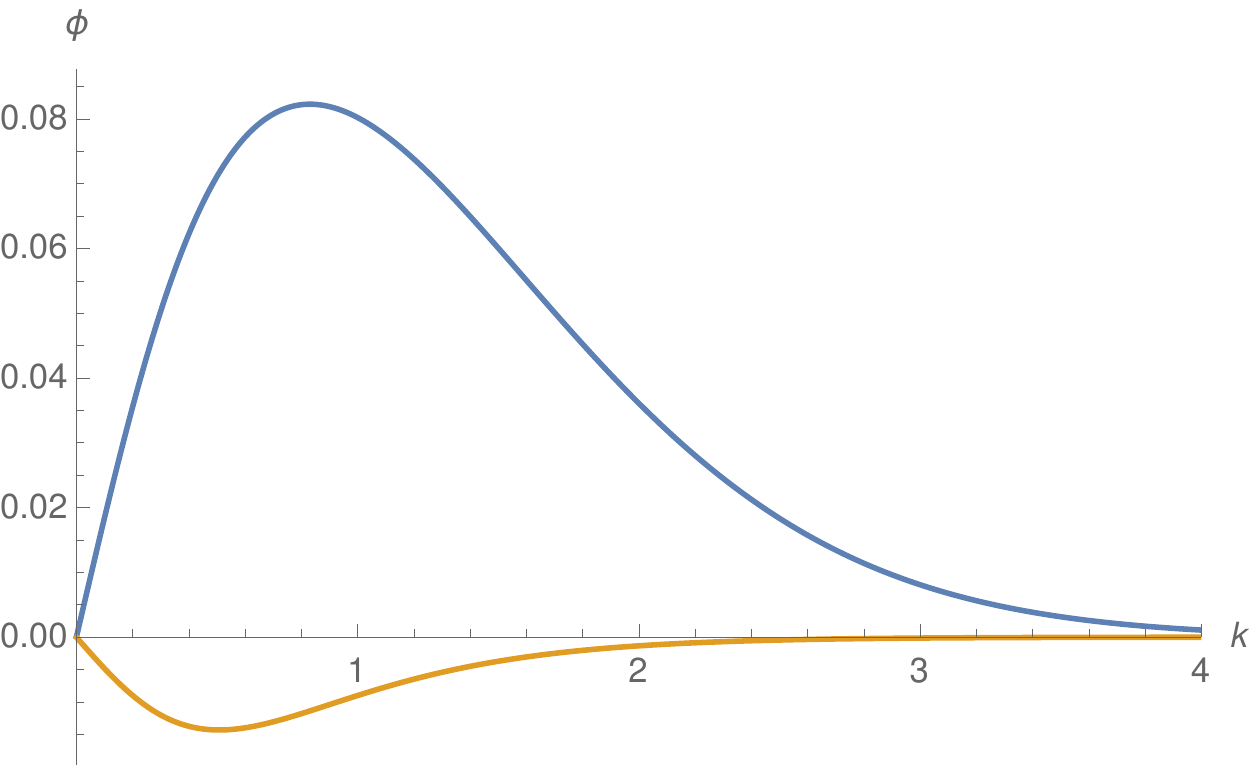} \hspace{1pt}
\includegraphics[width=0.30\columnwidth]{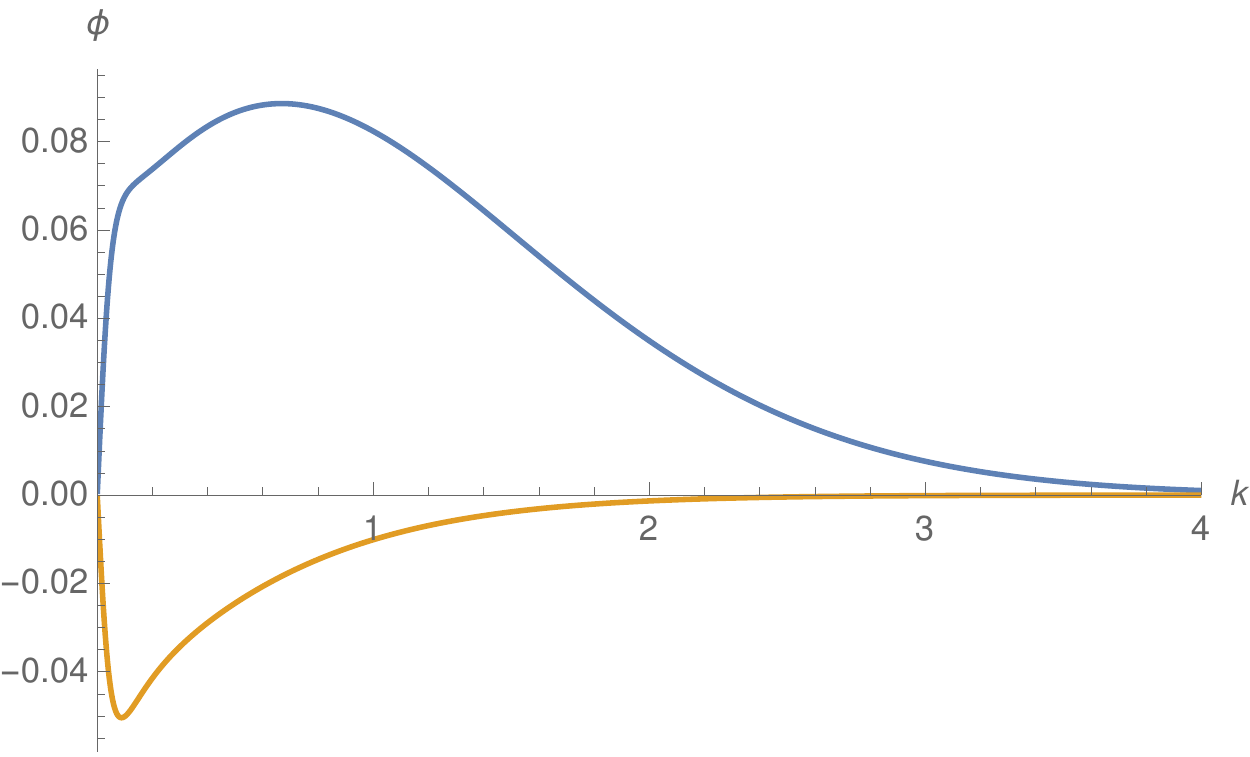} \hspace{1pt}
\includegraphics[width=0.30\columnwidth]{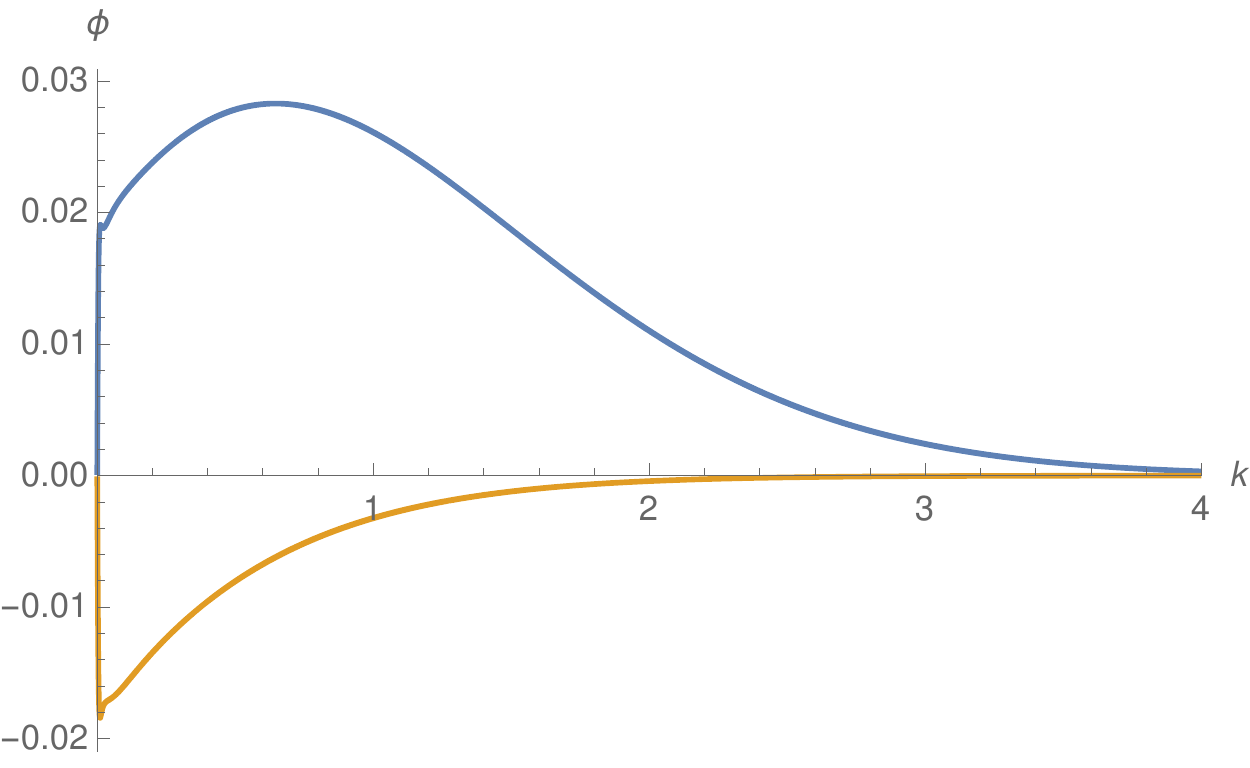}
\caption{The $J^{PC}=1^{++}$, $^3 P_1$ normalized axial vector (A) radial wave functions $\phi^+$ (in blue) and $\phi^-$ (in yellow), from left to right for the ground state vacuum and first two replicas,  in dimensionless in units of $K_0=1$. 
}\label{fig:axialvector2}
\end{figure}

Our bound state equations can be understood as an extension of the Shr\"odinger equation, since we now have positive $\phi^+$ and negative $\phi^-$ energy components of the wave functions, and this doubles the number of components. Nevertheless we have the same number 
of meson states as in the spectrum of the normal quark model. The mass splittings can also be related, as usual, to spin-tensor potentials.
 We use sparse matrices and matrix eigenvalues to compute the meson spectrum. 


We now address the main goal of this study: to show that the quark-antiquark bound states -- the mesons in the replicas -- all have real excess masses.

With the effective quark masses $m(k)$, we can compute the chiral angle, using ${ m(k) / k}=\tan[ \varphi (k) ]$.
According to the chiral theorem on the pion Salpeter amplitude \cite{Bicudo:1989si,Bicudo:2003fp}, the sine of the chiral angle $\sin[\varphi(k)]=m_c(k) /\sqrt{ k+ m_c(k)}$ should be proportional to the wave functions $\phi^{\pm}(k)$ of the pion. This is clearly the case, when we compare the mass represented in Fig. \ref{fig:masssolutionII} with the wavefunction shown in Fig. \ref{fig:pseudoscalar} .

For the ground state vacuum and for the first two replicas, in Figs. \ref{fig:pseudoscalar} and \ref{fig:scalar} , we show, respectively and in dimensionless in units of $K_0=1$, the wave functions $\phi^+$ and $\phi^-$, for the pseudoscalar meson and  the radial wave functions $\nu^+$ and $\nu^-$ for the scalar meson.  We notice that the number of nodes of these pseudoscalar wave functions depends on the replica where they are sitting, they have the same number of nodes as the constituent quark mass. This is expected due to the chiral theorem relating the pion wave function to the constituent quark mass \cite{Bicudo:1989si,Bicudo:2003fp} .

We also show, in Figs. \ref{fig:scalar} -- \ref{fig:axialvector2} , the radial wave functions of the scalar, vector and axial vector mesons. Notice how the wave functions change from one replica to another.
It is interesting to remark that, for the replicas, the excess masses of the mesons depend very little on the replica (and the physical vacuum) they are sitting in.

The masses scale with $K_0$, equivalent to the string tension. Even at finite temperature with a small $K_0$, this scenario will hold.


%
%
\section{Summary}

%
%
\begin{table}[h] 
\centering
\caption{
Masses of the mesons in the ground state vacuum (the true one) and excess masses$^{1,2}$ for the first two replicas, in units of $K_0$. For the mesons we show the pseudoscalar, the first excitation of the pseudoscalar, the scalar, the vector (mostly s-wave), the first excitation of the vector (mostly d-wave) and the two different axial vectors.
}
\label{tab:spectrum}
\begin{tabular}{c|c|c|c}
meson & in vacuum				& in replica 1$^1$ & in replica 2$^2$	\\ \hline
P \ $(J^{PC}=0^{-+})$ & 0.00 & 0.00 & 0.00 \\
P* $(J^{PC}=0^{-+})$ & 5.539 & 5.577 &  5.581 \\
S \ $(J^{PC}=0^{+ +})$ & 3.266 & 3.253 &  3.247 \\
V$_0$ $(J^{PC}=1^{--})$  & 2.686 & 2.823 & 2.836 \\
V$_2$ $(J^{PC}=1^{--})$ & 4.965 & 4.635 & 4.599 \\
A  \ $(J^{PC}=1^{+ -})$ & 4.103 & 3.784  & 3.723\\
A  \ $(J^{PC}=1^{+ +})$ & 4.665 & 4.602 & 4.596 \\
\end{tabular}
\end{table}

The real nature of the excess masses of the mesons, constitute the main result of this work. Our final results are shown in Table \ref{tab:spectrum}.  

Prior to this study, we were not sure what the masses of the mesons would be in the excited replica vacua. As in the false - chiral invariant - vacuum, we could possibly have tachyons. Their tachyonic free nature, show that the replicas and thus metastable. 

More recently, this study has been extended \cite{Bicudo:2021vrm} to the linear confining potential, with $\alpha=1$ (see Fig.\ref{fig:replicaslinear}) and to a small finite current mass $m_0$ (see Fig. \ref{fig:masssolution}) . 

As it stands, we cannot have all the low energy properties of hadronic
physics due to $S\chi SB$ without having replicated states as a
subproduct. It would  be extremely interesting to look wether  such  metastable
replicas  do actually exist in full QCD.

{\bf 
\noindent Acknowledgements}

P.B.\ and J.E.R.\ acknowledge the support of CeFEMA\mbox{} under the FCT contract for R\&D Units UIDB/04540/2020.



\bibliography{Bibliografia_1}

\end{document}